\documentclass[a4paper]{article}

\usepackage{INTERSPEECH2016}

\usepackage{graphicx}
\usepackage{amssymb,amsmath,bm}
\usepackage{textcomp}
\usepackage{epstopdf}
\usepackage{tabularx}
\usepackage{multirow}
\usepackage{hyperref}

\ninept

\title{Speech Enhancement In Multiple-Noise Conditions using Deep Neural Networks}

\makeatletter
\def\name#1{\gdef\@name{#1\\}}
\makeatother \name{{\em Anurag Kumar$^1$, Dinei Florencio$^2$}}

\address{$^1$Carnegie Mellon University, Pittsburgh, PA, USA - 15217 \\
 $^2$Microsoft Research, Redmond, WA USA - 98052\\
  {\small \tt alnu@andrew.cmu.edu, dinei@microsoft.com}
}

\begin{document}

  \maketitle
  %
\begin{abstract}
In this paper we consider the problem of speech enhancement in real-world like conditions where multiple noises can simultaneously corrupt speech. Most of the current literature on speech enhancement focus primarily on presence of single noise in corrupted speech which is far from real-world environments. Specifically, we deal with improving speech quality in office environment where multiple stationary as well as non-stationary noises can be simultaneously present in speech. We propose several strategies based on Deep Neural Networks (DNN) for speech enhancement in these scenarios. We also investigate a DNN training strategy based on psychoacoustic models from speech coding for enhancement of noisy speech. 
\end{abstract}
\vspace{-0.1in}
\section{Introduction}
\label{sec:intro}
\vspace{-0.05in}
Speech Enhancement (SE) is an important research problem in audio signal processing. The goal is to improve the quality and intelligibility of speech signals corrupted by noise. Due to its application in several areas such as automatic speech recognition, mobile communication, hearing aids etc. it has been an actively researched and several methods have been proposed over the past several decades \cite{loizou}\cite{cohen}.

The simplest method to remove additive noise by subtracting an estimate of noise spectrum from noisy speech spectrum was proposed back in 1979 by Boll \cite{boll}. The wiener filtering \cite{lim} based approach was proposed in the same year. MMSE estimator \cite {ephraim1} which performs non-linear estimation of short time spectral amplitude (STSA) of speech signal is another important work. A superior version MMSE estimation referred as Log-MMSE tries to minimize the mean square-error in the log-spectral domain \cite{ephraim2}. Other popular classic methods are called signal-subspace based methods \cite{ephraim3} \cite{hu}.

In recent years deep neural network (DNN) based learning architectures have been found to be very successful in related areas such as speech recognition \cite{hinton}\cite{dahl}\cite{graves1}\cite{graves2}. The success of deep neural networks (DNNs) in automatic speech recognition led to investigation of deep neural networks for noise suppression for ASR \cite{maas} and speech enhancement \cite{lu1} \cite{xu1} \cite{lu2}. The central theme in using DNNs for speech enhancement is that corruption of speech by noises is a complex process and a complex non-linear model like DNN is well suited for modeling it \cite{xu2}\cite{xia}. Although, there are very few exhaustive works on utility of DNNs for speech enhancement, it has shown promising results and can outperform classical SE methods. A common aspect in several of these works \cite{lu1}\cite{xia}\cite{lu2}\cite{tseng}\cite{xu1} is evaluation on matching or seen noise conditions. Matching or seen conditions implies the test \emph{noise types (e.g crowd noise)} are same as training. Unlike classical methods which are motivated by signal processing aspects, DNN based methods are \emph{data driven} approaches where matched noise conditions might not be ideal for evaluating DNNs for speech enhancement. In fact in several cases, the ``noise data set" used to create the noisy test utterances is same as the one used in training. This results in high similarity (same) between the training and test noises where it is not hard to expect that DNN would outperform other methods. Thus, a more thorough analysis even in matched conditions needs to be done by using variations of the selected noise types which have not been used during training. 

Unseen or mismatched noise conditions refer to the situations when the model (\emph{e.g DNN}) has not seen the test \emph{noise types} during training. For unseen noise conditions and enhancement using DNNs \cite{xu2} is a notable work. \cite{xu2} trains the network on a large variety of noise types and show that significant improvements can be achieved in mismatched noise conditions by exposing the network to large number of noise types. In \cite{xu2} ``noise data set" used to create the noisy test utterances is disjoint from that used during training although some of the test \emph{noise types} such as \emph{Car, Exhibition} would be similar to a few training noise types such as \emph{Traffic and Car Noise, Crowd Noise}. Some post-processing strategies were also used in this work to obtain improvements. Although, unseen noise conditions present a relatively difficult scenario compared to the seen one, it is still far from real-world applications of speech enhancement. In real-world we expect the model to not only perform equally well on large variety of noise types (seen or unseen) but also on non-stationary noises. More importantly, speech signals are usually corrupted by multiple noises of different types in real world situations and hence removal of single noise signals as done in all of the previous works is restrictive. In environments around us, multiple noises occur simultaneously with speech. This multiple \emph{noise types} conditions are clearly much harder and complex to remove or suppress. To analyze and study speech enhancement in these complex situations we propose to move to an \emph{environment specific} paradigm. In this paper we focus on \emph{office-environment} noises and propose different methods based on DNNs for speech enhancement in office-environment. We collect large number of \emph{office-environment} noises and in any given utterance several of these noises can be simultaneously present along with speech (details of dataset in later sections). We also show that \emph{noise-aware training} \cite{seltzer} proposed for noise robust speech recognition are helpful in speech enhancement as well in these complex noise conditions. We specifically propose to use running noise estimate cues, instead of stationary noise cues used in \cite{seltzer}. We also propose and evaluate strategies combining DNN and psychoacoustic models for speech enhancement. The main idea in this case is to change the error term in DNN training to address frequency bins which might be more important for speech enhancement. The criterion for deciding importance of frequency are derived from psychoacoustic principles. 

Section 2 describes the basic problem and different strategies for training DNNs for speech enhancement in multiple noise conditions, Section 3 first gives a description of datasets, experiments and results. We conclude in Section 4.
\vspace{-0.10in}
\section{DNN based Speech Enhancement}
\vspace{-0.05in}
Our goal is speech enhancement in conditions where multiple-noises of possibly different types might be simultaneously corrupting the speech signal. Both stationary and non-stationary noises of completely different acoustic characteristics can be present. This multiple-mixed noise conditions are close to real world environments. Speech corruption under these conditions are much more complex process compared to that by single noise and hence enhancement becomes a harder task. DNNs with their high non-linear modeling capabilities is presented here for speech enhancement in these complex situations. 

Before going into actual DNN description, the target domain for neural network processing needs to be specified first. Mel-frequency spectrum \cite{lu1}\cite{lu2}, ideal binary mask, ideal ratio mask, short-time Fourier transform magnitude and its mask \cite{wang}\cite{narayana}, log-power spectra are all potential candidates. \cite{xu2} showed that log-power spectra works better than other targets and we work in log-power spectra domain as well. Thus, our training data consists of pairs of log-power spectra of noisy and the corresponding clean utterance. We will simply refer to the log-power spectra as \emph{feature} for brevity at several places.

Our DNN architecture is a multilayer feed-forward network. The input to the network are the noisy feature frames and the desired output is the corresponding clean feature frames. Let $\mathbf{N}(t,f)=log(|STFT(n^u)|^2)$ be the log-power spectra of a noisy utterance $n^u$ where $STFT$ is the short-time Fourier transform. $t$ and $f$ represent time and frequency respectively and $f$ goes from $0$ to $N=(DFT\,\,size)/2$. Let $\mathbf{n}_t$ be the $t^{th}$ frame of $\mathbf{N}(t,f)$ and the context-expanded frame at $t$ be represented as $\mathbf{y}_t$, where $\mathbf{y}_t$ is given by
\begin{equation}
\label{contx}
\mathbf{y}_t = [\mathbf{n}_{t-\tau},....,\mathbf{n}_{t-1},\mathbf{n}_{t},\mathbf{n}_{t+1},....\mathbf{n}_{t+\tau}]
\vspace{-0.07in}
\end{equation}
Let $\mathbf{S}(t,f)$ be the log-power spectra of clean utterance corresponding to $n^u$. The $t^{th}$ clean feature frame from $\mathbf{S}(t,f)$ corresponds to $\mathbf{n}_t$ and is denoted as $\mathbf{s}_t$. We train our network with multi-condition speech \cite{seltzer} meaning the input to the network is $\mathbf{y}_t$ and the corresponding desired output is $\mathbf{s}_t$. The network is trained using back-propagation algorithm with mean-square error (MSE) (Eq. \ref{mse}) as error-criterion. Stochastic gradient descent over a minibatch is used to update the network parameters. 
\vspace{-0.1in}
\begin{equation}
\label{mse}
MSE = \frac{1}{K} \sum_{k=1}^{K} ||\mathbf{\hat{s}}_t - \mathbf{s}_t||^2 + \lambda ||\mathbf{W}||_{2}^2
\vspace{-0.1in}
\end{equation}
In Eq. \ref{mse}, $K$ is the size of minibatch and $\mathbf{\hat{s}}_t = f(\Theta, \mathbf{y}_t)$ is the output of the network. $f(\Theta)$ represents the highly non-linear mapping performed by the network. $\Theta$ collectively represents the weights ($\mathbf{W}$) and bias ($\mathbf{b}$) parameters of all layers in the network. The term $\lambda ||\mathbf{W}||_{2}^2$ is regularization term to avoid overfitting during training. A common thread in almost all of the current works on neural network based speech enhancement such as \cite{lu1} \cite{lu2} \cite{xia}\cite{xu2} is the use of either RBM or autoencoder based pretraining for learning network. However, given sufficiently large and varied dataset the pretraining stage can be eliminated and in this paper we use random initialization to initialize our networks. 

Once the network has been trained it can be used to obtain an estimate of clean log-power spectra for a given noisy test utterance. The STFT is then obtained from the log-power spectra. The STFT along with phase from noisy utterance is used to reconstruct the time domain signal using the method described in \cite{griffin}.
\vspace{-0.1in}
\subsection{Feature Expansion at Input}
\vspace{-0.05in}
We expand the feature at input by two methods both of which are based on the fact that feeding information about noise present in the utterance to the DNN is beneficial for \emph{speech recognition} \cite{seltzer}. \cite{seltzer} called it \emph{noise-aware training} of DNN. The idea is that the non-linear relationship between noisy-speech log-spectra, clean-speech log-spectra and the noise log-spectra can be modeled by the non-linear layers of DNN by directly giving the noise log-spectra as input to the network. This is simply done by augmenting the input to the network $\mathbf{y}_t$ with an estimate of the noise ($\mathbf{\hat{e}}_t$) in the frame $\mathbf{n}_t$. Thus the new input to the network becomes
\begin{equation}
\label{featexp}
\mathbf{y'}_t = [\mathbf{n}_{t-\tau},..,\mathbf{n}_{t-1},\mathbf{n}_{t},\mathbf{n}_{t+1},..\mathbf{n}_{t+\tau},\mathbf{\hat{e}}_t]
\vspace{-0.1in}
\end{equation}
The same idea can be extended to speech enhancement as well. \cite{seltzer} used stationary noise assumption and in this case the $\mathbf{\hat{e}}_t$ for the whole utterance is fixed and obtained using the first few frames ($F$) of noisy log-spectra
\vspace{-0.15in}
\begin{equation}
\label{staf}
\mathbf{\hat{e}}_t = \mathbf{\hat{e}} = \frac{1}{F} \sum_{t=1}^{F} \mathbf{n}_{t}
\vspace{-0.10in}
\end{equation}
However, under our conditions where multiple noises each of which can be non-stationary, a \emph{running estimate of noise} in each frame might be more beneficial. We use the algorithm described in \cite{gerkmann} to estimate $\mathbf{\hat{e}}_t$ in each frame and use it in Eq \ref{featexp} for input feature expansion. We expect the running estimate to perform better compared to Eq. \ref{staf} in situations where noise is dominant (low SNR) and noise is highly non-stationary.
\vspace{-0.1in}
\subsection{Psychoacoustic Models based DNN training}
\vspace{-0.05in}
The sum of squared error for a frame (SE, Eq \ref{sqe}) used in Eq. \ref{mse} gives equal importance to the error at all frequency bins. This means that all frequency bins contribute with equal importance in gradient based parameter updates of network. However, for intelligibility and quality of speech it is well known from pyschoacoustics and audio coding \cite{painter}\cite{fletcher}\cite{greenwood}\cite{zwislocki}\cite{scharfl}\cite{hellman} that all frequencies are not equally important. Hence, the DNN should focus more on frequencies which are more important. This implies that the same error for different frequencies should contribute in accordance of their importance to the network parameter updates. We achieve this by using weighted squared error (WSE) as defined in Eq \ref{wsqe}
\vspace{-0.1in}
\belowdisplayskip=0pt
\begin{align}
\quad\quad\quad SE  = & ||\mathbf{\hat{s}}_t - \mathbf{s}_t||^2_2 =\sum_{i=0}^N (\hat{s}^{i}_t - s^{i}_t)^2  \label{sqe}\\ 
WSE = & ||\mathbf{w}_t \odot (\mathbf{\hat{s}}_t - \mathbf{s}_t)||^{2}_2 = \sum_{i=0}^N (w^{i}_t)^2 (\hat{s}^{i}_t - s^{i}_t)^2 \label{wsqe}
\end{align}
$\mathbf{w}_t > \mathbf{0}$ is the weight vector representing the \emph{frequency-importance} pattern for the frame $\mathbf{s}_t$ and $\odot$ represents the element wise product. The DNN training remains same as before except that the gradients are now computed with respect to the new mean weighted squared error (MWSE, Eq \ref{mwse}) over a minibatch.
\vspace{-0.15in}
\begin{equation}
\label{mwse}
MWSE = \frac{1}{K} \sum_{k=1}^{K} ||\mathbf{w}_t \odot (\mathbf{\hat{s}}_t - \mathbf{s}_t)||^{2}_2 + \lambda ||\mathbf{W}||_{2}^2
\end{equation}
\vspace{-0.02in}
The bigger question of describing the \emph{frequency-importance} weights needs to be answered. We propose to use psychoacoustic principles frequently used in audio coding for defining $\mathbf{w}_t$ \cite{painter}. Several psychoacoustic models characterizing human audio perception such as absolute threshold of hearing, critical frequency band and masking principles have been successfully used for efficient high quality audio coding. All of these models rely on the main idea that for a given signal it is possible to identify time-frequency regions which would be more important for human perception. We propose to use absolute threshold of hearing (ATH) \cite{fletcher} and masking principles \cite{painter}\cite{scharfl}\cite{hellman} to obtain our \emph{frequency-importance} weights. The ATH based weights leads to a global weighing scheme where the weight $\mathbf{w}_t=\mathbf{w}^g$ is same for the whole data. Masking based weights are frame dependent where $\mathbf{w}_t$ is obtained using $\mathbf{s}_t$.
\vspace{-0.10in}
\subsubsection{ATH based Frequency Weighing}
\label{gwts}
\vspace{-0.05in}
The ATH defines the minimum sound energy (sound pressure level in dB) required in a pure tone to be detectable in a quiet environment. The relationship between energy threshold and frequency in Hertz ($fq$) is approximated as \cite{terhardt}
\begin{equation}
\resizebox{1.0\columnwidth}{!}{$ATH(fq)=3.64(\frac{fq}{1000})^{-0.8}-6.5e^{-0.6(\frac{fq}{1000}-3.3)^2}+10^{-3}(\frac{fq}{1000})^4$}
\end{equation}
ATH can be used to define \emph{frequency-importance} because lower absolute hearing threshold implies that the corresponding frequency can be easily heard and hence more important for human perception. Hence, the \emph{frequency-importance} weights $\mathbf{w}^g$ can be defined to have an inverse relationship with $ATH(fq)$. We first compute the $ATH(fq)$ at center frequency of each frequency bin ($f=0\,\,to\,\,N$) and then shift all thresholds such that the minimum lies at 1. The weight $w^g_f$ for each $f$ is then the inverse of the corresponding shifted threshold. To avoid assigning a 0 weight ($ATH(0)=\infty$) to $f=0$ frequency bin the threshold for it is computed at $3/4th$ of the frequency range for $0^{th}$ frequency bin.
\vspace{-0.10in}
\subsubsection{Masking Based Frequency Weighing}
\vspace{-0.05in}
Masking in frequency domain is another psychoacoustic model which has been efficiently exploited in perceptual audio coding. Our idea behind using masking based weights is that noise will be masked and hence inaudible at frequencies where speech power is dominant.
More specifically, we compute a masking threshold $MTH(fq)$ based on a triangular spreading function with slopes of +25 and -10dB per bark computed over each frame of clean magnitude spectrum \cite{painter}. $MTH(fq)_t$ are then scaled to have maximum of $1$. The absolute values of logarithm of these scaled thresholds are then shifted to have minimum at $1$ to obtain $\mathbf{w}_t$. Note that, for simplicity, we ignore the differences between tone and noise masking. In all cases weights are normalized to have their square sum to $N$. 
\vspace{-0.10in}
\section{Experiments and Results}
\vspace{-0.05in}
As stated before our goal is to study speech enhancement using DNN in conditions similar to real-world environments. We chose \emph{office-environment} for our study. We collected a total of $95$ noise samples as representative of noises often observed in office environments. Some of these have been collected at Microsoft and the rest have been obtained mostly from \cite{fsound} and few from \cite{nonspeech}. We provide more details on these noises and their spectrograms on this link \cite{nsset}. We randomly select $70$ (set $NTr$) of these noises for creating noisy training data and the rest $25$ ( set $NTe$) for creating noisy testing data. 
Our clean database source is TIMIT\cite{timit}, from which train and test sets are used accordingly in our experiments. Our procedure for creating multiple-mixed noise situation is as follows. 

For a given clean speech utterance from TIMIT training set a random number of noise samples from $NTr$ are first chosen. This random number can be at most $4$ i.e at most four noises can be simultaneously present in the utterance. The chosen noise samples are then mixed and added to the clean utterance at a random SNR chosen uniformly from $-5\,dB$ to $20\,dB$. All noise sources receive equal weights. This process is repeated several times for all utterances in the TIMIT training set till the desired amount of training data have been obtained. For our testing data we randomly choose $250$ clean utterances from TIMIT test set and then add noise in a similar way. The difference now is that the noises to be added are chosen from $NTe$ and the SNR values for corruption in test case are fixed at $\{-5,0,5,10,15,20\}$ dBs. This is done to obtain insights into performance at different degradation levels. A validation set similar to the test set is also created using another $250$ utterances randomly chosen from TIMIT test set. This set is used for model selection wherever needed. To show comparison with classical methods we use Log-MMSE as baseline.  

We first created a training dataset of approximately $25$ hours. Our test data consists of $1500$ utterances of about $1.5$ hours. Since, DNN is data driven approach we created another training dataset of about $100$ hours to study the gain obtained by $4$ fold increase in training data. All processing is done at 16KHz sampling rate with window size of $16ms$ and window is shifted by $8ms$. All of our DNNs consists of $3$ hidden layers with $2048$ nodes and sigmoidal non-linearity.  The values of $\tau$ and $\lambda$ are fixed throughout all experiments at $5$ and $10^{-5}$. The $F$ in Eq \ref{staf} is $8$. The learning rate is usually kept at $0.05$ for first $10$ epochs and then decreased to $0.01$ and the total number of epochs for DNN training is $40$. The best model across different epochs is selected using the validation set. CNTK\cite{yu} is used for all of our experiments.

We measure both speech quality and speech intelligibility of the reconstructed speech. PESQ \cite{pesq} is used to measure the speech quality and STOI \cite{stoi} to measure intelligibility. To directly substantiate the ability of DNN in modeling complex noisy log spectra to clean log-spectra we also measure speech distortion and noise reduction measure \cite{lu1}. Speech distortion basically measures the error between the DNN's output (log spectra) and corresponding desired output or target (clean log spectra). It is defined for an utterance as $\text{SD}=\frac{\sum_{t=1}^T |\mathbf{\hat{s}}_t - \mathbf{s}_t|}{T}$. Noise reduction measures the reduction of noise in each noisy-feature frame $\mathbf{n}_t$ and is defined as $\text{NR}=\frac{\sum_{t=1}^T |\mathbf{\hat{s}}_t - \mathbf{n}_t|}{T}$. Higher NR implies better results, however very high NR might result in higher distortion of speech. This is not desriable as SD should be as low as possible. We will be reporting mean over all utterances for all four measures. 

Table \ref{tab:res} shows the PESQ measurement averaged over all utterances for different cases with $25$ hours of training data. In Table \ref{tab:res} LM represents results for Log-MMSE, BD for DNN without feature expansion at input. BSD is DNN with feature expansion at input ($\mathbf{y'}_t$)  using Eq \ref{staf} and BED DNN with $\mathbf{y'}_t$ using a running estimate of noise ($\mathbf{\hat{e}}_t$) in each frame using \cite{gerkmann}. Its clear that DNN based speech enhancement is much superior compared to Log-MMSE for speech enhancement in multiple-noise conditions. DNN results in significant gain in PESQ at all SNRs. The best results are  obtained with BED. At lower SNRs ($-5$, $0$ and $5$ dB) the absolute mean improvement over noisy PESQ is $0.43$, $0.53$ and $0.60$ respectively which is about $30\%$ increment in each case.  At higher SNRs the average improvement is close to $20\%$. Our general observation is that DNN with weighted error training (MWSE) leads to improvement over their respective non-weighted case only at very low SNR values. Due to space constraints we show results for one such case, BSWD, which corresponds to weighted error training of BSD. The better of the two weighing schemes is presented. On an average we observe that improvement exist only for $-5$dB. 
\begin{table}[t]
\caption{Avg. PESQ results for different cases}
\resizebox*{1.0\columnwidth}{!}{
\begin{tabular}{|c|c|c|c|c|c|c|}
\hline  
SNR(dB)&Noisy&LM&BD&BSD&BED&BSWD \\
\hline 
-5&1.46&1.61&1.85&1.84&1.89&1.88\\
\hline
0&1.77&2.02&2.26&2.28&2.30&2.26 \\
\hline
5&2.11&2.41&2.64&2.70&2.71&2.65\\
\hline
10&2.53&2.83&3.05&3.12&3.12&3.04\\
\hline
15&2.88&3.14&3.37&3.43&3.42&3.42\\
\hline
20&3.23&3.44&3.61&3.68&3.68&3.60\\
\hline
\end{tabular}
}
\label{tab:res}
\vspace{-0.15in}
\end{table}

\begin{figure}[t]
\centering
\includegraphics[width=1.0\columnwidth,height=1.2in]{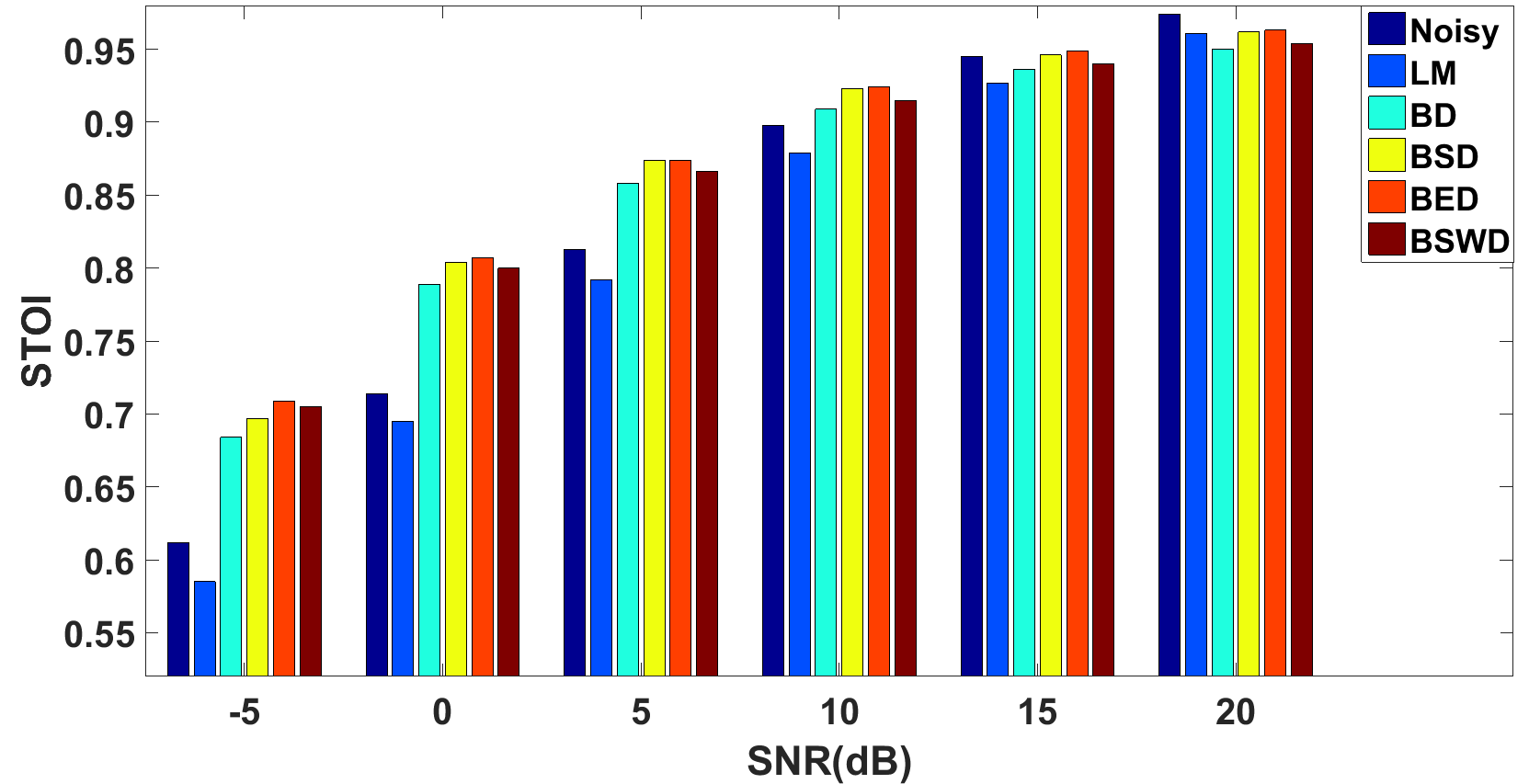}
\vspace{-0.25in}
\caption{Average STOI Comparison for Different Cases}
\label{fig:stoi}
\vspace{-0.2in}
\end{figure}

For real world applications its important to analyze the intelligibility of speech along with speech quality. STOI is one of the best way to objectively measure speech intelligibility\cite{stoi}. It ranges from $0$ to $1$ with higher score implying better intelligibility. Figure \ref{fig:stoi} shows speech intelligibility for different cases. We observe that in our multiple-noise conditions, although speech quality (PESQ) is improved by Log-MMSE it is not the case for intelligibility(STOI). For Log-MMSE, STOI is reduced especially at low SNRs where noise dominate. On the other hand, we observe that DNN results in substantial gain in STOI at low SNRs. BED again outperforms all other methods where $15-16\%$ improvement in STOI over noisy speech is observed at $-5$ and $0$ dB. 
\begin{table}[t]
\label{tab:25sdnr}
\centering
\caption{Average SD and NR for different cases}
\resizebox*{1.0\columnwidth}{!}{
\begin{tabular}{|c|c|c|c|c|c|c|c|c|c|c|}
\hline  
SNR &\multicolumn{2}{c|}{LM} &  \multicolumn{2}{c|}{BD} & \multicolumn{2}{c|}{BSD} & \multicolumn{2}{c|}{BED} & \multicolumn{2}{c|}{BSWD}\\ 
\cline{2-11}
dB&NR&SD&NR&SD&NR&SD&NR&SD&NR&SD\\
\hline 
-5 & 3.18&3.11&4.12&2.02&4.55&1.90&4.10&1.93&4.19&1.92\\
\hline
0 & 3.07&2.72&3.47&1.75&3.78&1.63&3.51&1.65&3.56&1.66\\
\hline
5 & 2.88&2.37&2.9&1.48&3.04&1.39&2.88&1.4&2.96&1.41\\
\hline
10 & 2.53&2.02&2.32&1.27&2.36&1.18&2.25&1.19&2.34&1.19\\
\hline
15 & 2.16&1.74&1.81&1.12&1.79&1.03&1.71&1.03&1.79&1.04\\
\hline
20 & 1.81&1.48&1.41&1.00&1.35&0.89&1.28&0.89&1.36&0.91\\
\hline
\end{tabular}
}
\vspace{-0.05in}
\end{table}

\begin{figure}[t]
\centering
\includegraphics[width=1.0\columnwidth,height=2.0in]{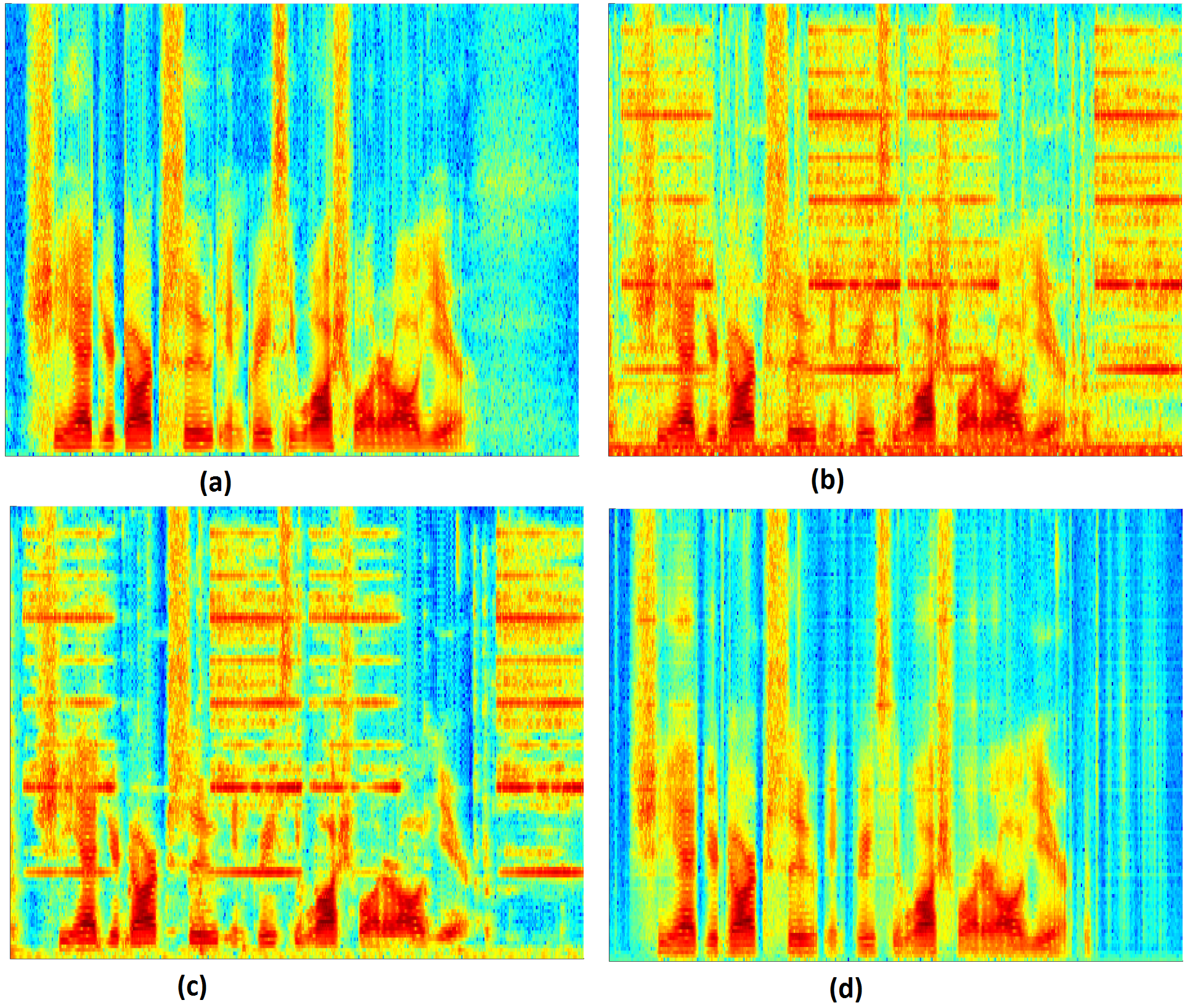}
\vspace{-0.2in}
\caption{Spectrograms (a) clean utterance (b) Noisy (c) Log-MMSE (d) DNN Enhancement (BED)}
\label{fig:specs}
\vspace{-0.20in}
\end{figure}

For visual comparison, spectrograms for an utterance corrupted at 5dB SNR with highly non-stationary multiple noises (\emph{printer} and \emph{typewriter} noises along with \emph{office-ambiance} noise) is shown in Figure \ref{fig:specs}. The PESQ values for this utterance  are; \emph{noisy} = $2.42$, Log-MMSE = $2.41$,  BED DNN = $3.10$. The audio files corresponding to Figure \ref{fig:specs} have been submitted as additional material. Clearly, BED is far superior to Log-MMSE which completely fails in this case. For BEWD (not shown due to space constraint) the PESQ obtained is $3.20$ which is highest among all methods. This is observed for several other test cases where the weighted training leads to improvement over corresponding non-weighted case; although on average we saw previously that it is helpful only at very low SNR($-5$ dB). This suggests that weighted DNN training might give superior results by using methods such as dropout \cite{hinton2012} which helps in network generalization. \textbf{Some more audio and spectrogram examples are available at \cite{nsset}}.

The SD and NR values for different DNN's are shown in Table \ref{tab:25sdnr}. For the purpose of comparison we also include these values for Log-MMSE. We observe that in general DNN architectures compared to LM leads to increment in noise reduction and  decrease in speech distortion which are the desirable situations. Trade-off between SD and NR exists and the optimal values leading to improvements in measures such as PESQ and STOI varies with test cases.

Finally, we show the PESQ and STOI values on test data for DNNs trained with $100$ hours of training data in Table \ref{tab:100pesqstoi}. Larger training data clearly leads to a more robust DNN leading to improvements in both PESQ and STOI. For all DNN models improvement over the corresponding $25$ hour training can be observed. 
\begin{table}[t!]
\label{tab:100pesqstoi}
\centering
\caption{Average PESQ and STOI using $100$ hours training data}
\resizebox*{1.0\columnwidth}{!}{
\begin{tabular}{|c|c|c|c|c|c|c|c|c|}
\hline  
SNR &\multicolumn{2}{c|}{Noisy} & \multicolumn{2}{c|}{BD} & \multicolumn{2}{c|}{BSD} & \multicolumn{2}{c|}{BED}\\ 
\cline{2-9}
dB&PESQ&STOI&PESQ&STOI&PESQ&STOI&PESQ&STOI\\
\hline 
-5 & 1.46&0.612&1.92&0.703&1.93&0.712&1.96&0.717\\
\hline
0 & 1.77&0.714&2.32&0.804&2.35&0.812&2.36&0.812\\
\hline
5 & 2.11&0.813&2.69&0.872&2.75&0.881&2.74&0.879\\
\hline
10 & 2.53&0.898&3.09&0.923&3.14&0.928&3.14&0.928\\
\hline
15 & 2.88&0.945&3.40&0.950&3.44&0.954&3.44&0.953\\
\hline
20 & 3.23&0.974&3.67&0.965&3.72&0.970&3.71&0.970\\
\hline
\end{tabular}
}
\vspace{-0.15in}
\end{table}
\vspace{-0.15in}
\section{Conclusions}
\label{sec:conclusions}
\vspace{-0.05in}
In this paper we studied speech enhancement in complex conditions which are close to real-word environments. We analyzed effectiveness of deep neural network architectures for speech enhancement in multiple noise conditions; where each noise can be stationary or non-stationary. Our results show that DNN based strategies for speech enhancement in these complex situations can work remarkably well. Our best model gives an average PESQ increment of $23.97\%$ across all test SNRs. At lower SNRs this number is close to $30\%$. This is much superior to classical methods such as Log-MMSE. We also showed that augmenting noise cues to the network definitely helps in enhancement. We also proposed to use running estimate of noise in each frame for augmentation, which turned out to be especially beneficial at low SNRs. This is expected as several of the noises in the test set are highly non-stationary and at low SNRs these dominant noises should be estimated in each frame. We also proposed psychoacoustics based weighted error training of DNN. Our current experiments suggests that it is helpful mainly at very low SNR. However, analysis of several test cases suggests that network parameter tuning and dropout training which improves generalization might show the effectiveness of weighted error training. We plan to do a more exhaustive study in future. However, this work does give conclusive evidence that DNN based speech enhancement can work in complex multiple-noise conditions like in real-world environments.       
\newpage
\eightpt
\bibliographystyle{IEEEtran}
\bibliography{refs2}


\end{document}